\documentstyle[amssymb,aps,prl,twocolumn,epsf]{revtex}

\begin{document}
\draft

\flushbottom
\twocolumn[\hsize\textwidth\columnwidth\hsize\csname
@twocolumnfalse\endcsname
\title{Droplets in Disordered Metallic Quantum Critical Systems}
\author{A.~H.~Castro Neto$^{a,*}$ and B.~A.~Jones $^b$}
\address
{$^{a}$Department of Physics, Boston University, Boston, MA 02215 \\
$^b$ IBM Almaden Research Center, San Jose, CA 95120-6099}
\date{\today }
\maketitle
\widetext\leftskip=1.9cm\rightskip=1.9cm\nointerlineskip\small
\begin{abstract}
\hspace*{2mm}
We present a field theory for a structurally disordered magnetic
system coupled to a metallic environment near a quantum critical point.
We show that close to the magnetic quantum critical point droplets are formed
due to the disorder and undergo dissipative quantum dynamics. We
show that the problem has a characteristic energy scale, the droplet Kondo
temperature, that determines the crossover energy scale 
from weak to strong coupling.
Our results have direct significance
for the Griffiths-McCoy singularities of itinerant magnets.
\end{abstract}
] \narrowtext

The behavior of itinerant magnets close to quantum critical
points (QCP) has been a subject of intense research in the last few
years. It has been found experimentally that a broad class of systems
show anomalous metallic behavior in the paramagnetic phase \cite{review}.
Since a large number of such systems are structurally disordered due
to chemical substitution the question arises of the importance
of disorder for the understanding of the anomalies observed
in the experiments. We have proposed recently that the anomalous
behavior observed in some of these systems can be understood in
terms of Griffiths-McCoy singularities close to a QCP \cite{us1,us2}.
These singularities occur in the context of percolation theory on
a discrete lattice when clusters of spins
tunnel quantum mechanically. In the presence of a metallic environment
we have shown that electrons scatter against clusters leading to a
{\it cluster Kondo temperature} associated with their dissipative
quantum dynamics \cite{us2}. We found that while dissipation freezes
the clusters when the system is close enough to the QCP, there is
still possibility of quantum behavior in a large region in the
parameter space around the QCP. In this case small Kondo temperatures
can be obtained even for relatively small clusters dropping the
requirement of cluster ``rarity'' as a condition for anomalous magnetic
behavior \cite{us2}. These results put our approach in proximity to
the single ion Kondo disorder theories \cite{miranda}.

In a recent paper Millis, Morr and Schmalian (MMS) \cite{mms}
proposed that a single isolated local perturbation close to a QCP of an
itinerant Ising magnet produces large droplets with internal structure
that are blocked from tunneling due to dissipative effects.
While that theory has similar features to the Griffiths-McCoy scenario
proposed by us, their theory is supposed to be valid for clean
critical systems
with a vanishing small amount of local defects. Experimentally this
scenario can be realized by introducing a very small amount of impurities
into clean stoichiometric compounds like CeRu$_2$Ge$_2$ \cite{jaccard}
or CePd$_2$Si$_2$ \cite{lonzarich} and driving the system close to a
QCP by application of hydrostatic pressure.

In this work we study the problem of a disordered alloy such as
U$_{1-x}$Y$_x$Pd$_3$ \cite{maple} or UCu$_{5-x}$Pd$_x$ \cite{ucupd}
with a finite density of defects or impurities and where the
distance from the QCP is controlled by chemical substitution. We show that,
for the same field theory studied by MMS, a {\it finite density} of
impurities leads to Griffiths-McCoy singularities in a region close to
the QCP. We show that in the presence
of disorder the droplet is structureless and its average size is set
by the magnetic correlation length, $\xi$. More importantly,
we demonstrate that the droplets have a finite droplet Kondo
temperature, $T_K$, that varies continuously with the distance from the
QCP and vanishes when the system is sufficiently close to it.
These results are in agreement with
our previous {\it microscopic} analysis of
the disordered Kondo lattice model \cite{us2} but
also apply to other models of strongly correlated electrons
such as the Hubbard model \cite{hertz}.

The starting point of our analysis of droplet formation close to
a QCP is the Hertz action for a critical itinerant magnetic
system in $d$ spatial dimensions \cite{hertz} (we use units such that
$\hbar=k_B=1$):
\begin{eqnarray}
S = \frac{1}{2 \beta} \sum_{n,{\bf q}} \left(\omega_n^2 + \gamma \Gamma_q
|\omega_n| + q^2 + r \right) |\varphi(\omega_n,{\bf q})|^2
\nonumber
\end{eqnarray}
where $\omega_n$ and ${\bf q}$ are the Matsubara frequency and momentum,
respectively, $\gamma$ is the coupling between the order parameter
$\varphi({\bf x},\tau)$ and the particle-hole continuum, $\Gamma_q$
gives the momentum dependence of the dissipative coupling
($\Gamma_q = q^{-\zeta}$ with $\zeta=0$ for antiferromagnets, $\zeta=1$ for
clean ferromagnets and $\zeta=2$ for disordered ferromagnets).
The distance from the QCP is controlled by $r$. In the ordered phase
$r<0$ indicating an instability towards long range order and at the QCP
we have $r=0$. In this work we focus entirely in the paramagnetic phase
where $r>0$ so that we can parametrize $r=\xi^{-2}$.
Disorder is introduced into
the problem as a random variation of $r$ in real space:
\begin{eqnarray}
S_{dis} = -\frac{1}{2}\int d{\bf x} \int_0^{\beta} d\tau \, \delta r({\bf x})
\, \, \varphi^2({\bf x},\tau)
\nonumber
\end{eqnarray}
which is assumed to be Gaussian distributed with width $u$, that is,
the probability distribution is given by:
\begin{eqnarray}
P(\delta r) \propto \exp\left\{-\frac{1}{4 u} \int d{\bf x} \, \, 
(\delta r({\bf x}))^2 \right\}
\label{probr}
\end{eqnarray}
so that $\overline{\delta r({\bf x})} =0$ and
\begin{eqnarray}
\overline{\delta r({\bf x}) \delta r({\bf y})} =
u \, \, \delta^d({\bf x}-{\bf y})
\nonumber
\end{eqnarray}
where the average is calculated with (\ref{probr}). The reason for
the appearence of droplets in the problem is that the Gaussian
distribution (\ref{probr}) allows for local values of $\delta r({\bf x})$
(the tails of the distribution) such that $r-\delta r({\bf x}) < 0$,
that is, it allows for local order even in the absence of long range order.
These locally ordered regions in a surrounding paramagnetic media are
called droplets.

Since we are not interested in one particular realization of the disorder
we study the average free energy using replicas \cite{replicas}.
We introduce $n$ replicas of the order parameter $\varphi_a$ with $a=1,...,n$
and calculate the average free energy, $\overline{F} = (\overline{Z^n}-1)/n$,
taking the limit of $n \to 0$ at the end of the calculation. It can
be easily shown that a new term is generated in the quantum action:
\begin{eqnarray}
{\cal S}_{dis} &=& - \frac{u}{4} \sum_{a,b} \int d{\bf x}
\int_0^{\beta} d\tau \int_0^{\beta} d\tau'
\varphi_a^2({\bf x},\tau) \varphi_b^2({\bf x},\tau') \, .
\nonumber
\end{eqnarray}
Notice that the disorder not only generates interactions between
fields in different replicas but also couples the fields in the
imaginary time direction. Our calculations can also be carried out
with a non-linear term of the form $g \varphi^4({\bf x},\tau)$
with no fundamental change in the results. Thus, in order to keep
the discussion simple we have dropped this term.

We first study the problem of droplet formation by investigating
the static, classical, part of the action.
In this case $\varphi_a({\bf x},\tau) =
\psi_a({\bf x})$ where $\psi_a$ is obtained from the variational
solution of the static part of the action:
\begin{eqnarray}
-\nabla^2 \psi_a({\bf x}) + r \psi_a({\bf x})
- u \beta \psi_a({\bf x}) \sum_b \psi^2_b({\bf x})
= 0 \, .
\label{replicaequation}
\end{eqnarray}
This equation is quite revealing. If the problem were to be replica
symmetric, that is, $\psi_a({\bf x}) = \psi_0({\bf x})$ for all values
of $a$, the last term in (\ref{replicaequation}) would scale like
$n \psi^2_0({\bf x})$ and would vanish in the limit of $n \to 0$.
This would imply that the only solution in the paramagnetic phase ($r>0$)
is the trivial solution $\psi_0=0$. Thus, in order for droplets
to form in the paramagnetic phase one needs the replica symmetry to be
broken. Here we follow Dotsenko \cite{dotsenko} who studied the
classical problem in detail and assume a non-trivial replica solution
such that for $a=1,...,k$ we have $\psi_a({\bf x}) = \psi_k({\bf x})$
while $\psi_a({\bf x})=0$ for $a=k+1,...,n$. Here $k>1$ is an integer
that determines the degree of the symmetry breaking process.
Using this particular solution we see that (\ref{replicaequation})
can be rewritten as:
\begin{eqnarray}
-\nabla^2 \psi_k({\bf x}) + r \psi_k({\bf x})
- u \beta k \psi_k^3({\bf x}) = 0 \, .
\label{repbrok}
\end{eqnarray}
This equation is non-linear Schr\"odinger equation
and can be thought as the equation for a classical particle moving
in the potential shown in Fig.\ref{pot}.

\begin{figure}
\epsfysize5 cm
\hspace{0cm}
\epsfbox{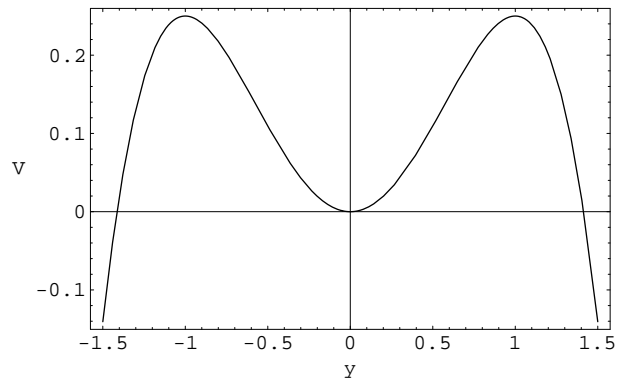}
\caption{Replica potential.}
\label{pot}
\end{figure}

A naive conclusion from this discussion would be that the
``energy'' would be minimized for $\psi = \pm \infty$, a solution
that is clearly unphysical. It should be kept in mind that
the replicated action is {\bf not} the actual free energy of
the problem; the latter is only obtained after we take the limit
of $n \to 0$ at the end of the calculation (this fact
is a standard consequence of the application of the replica method
\cite{replicas}). Dotsenko
has shown that the stable static solutions of this problem are
the maxima of Fig.\ref{pot}. Mathematically this can be proved by showing
that all the eigenvalues of the fluctuation matrix (the Hessian)
are real and positive \cite{dotsenko}.
The maxima are associated with the two
possible configurations of the droplet (spin configurations
pointing up and down).
Equation (\ref{repbrok}) can be rescaled if we define
\begin{eqnarray}
\psi_k(x) = \sqrt{\frac{r}{u \beta k}} \phi(\sqrt{r} x)
\label{psix}
\end{eqnarray}
so that $\phi(z)$ obeys a scale independent equation:
\begin{eqnarray}
-\nabla^2_z \phi(z) + \phi - \phi^3 = 0
\label{homo}
\end{eqnarray}
where $z = \sqrt{r} x$. The proper boundary conditions
are $\phi(0) =$ constant and $\phi(z \to \pm \infty) = 0$.
Equation (\ref{homo}) has exponentially
decaying solutions for $x \gg 1/\sqrt{r}$ and is smooth for
$x < 1/\sqrt{r}$. This can be contrasted with the solution
found by MMS for a local defect in which the
droplet is such that for $x<1/\sqrt{r}$ the droplet profile
decays like $1/x$. It is the $1/x$ decay that makes the MMS
problem special \cite{mms}. Here such a behavior does not occur in $d<4$.
Moreover, it is clear that the size of the droplet is the magnetic
correlation length:
\begin{eqnarray}
R \approx \frac{1}{\sqrt{r}} = \xi \, .
\nonumber
\end{eqnarray}
Thus droplets become arbitrarily large close
to the QCP.

The action for the static droplet can be calculated
by substitution of (\ref{psix})
into the original action, and the free energy, after summing
over the replicas, reads \cite{dotsenko}:
\begin{eqnarray}
\frac{F^0_D}{V} \approx - u r^{d/2}
\exp\left\{- \frac{r^{2-d/2} E_2}{4 u} \right\}
\,
\nonumber
\end{eqnarray}
where $E_N = \int d^d z \phi^{2 N}(z)$ and $V$ is the
volume of the system.
As pointed out by Dotsenko the non-analytic, non-perturbative,
dependence of the
free energy on the disorder strength $u$ for $d<4$ shows that the
static field theory reproduces the classical Griffiths
result in a percolating lattice \cite{griffiths}.
Obviously in order to study quantum or Griffiths-McCoy
singularities \cite{mccoy}
we have to allow the droplet to tunnel between the two
maxima of the potential in Fig.\ref{pot}.

In order to study the tunneling of the droplet we
assume a {\it rigid} droplet approximation \cite{mms}:
\begin{eqnarray}
\varphi({\bf x},\tau) = \psi_k({\bf x}) X(\tau)
\label{varphi}
\end{eqnarray}
where all the dynamics is encapsulated in $X(\tau)$.
This choice assumes that the droplet tunnels as a whole.
Direct substitution of (\ref{varphi}) into the action leads to:
\begin{eqnarray}
Z[X] \approx \int DX(\tau) \, \, e^{-r^{2-d/2} E_2 {\cal F}[X]/u}
\nonumber
\end{eqnarray}
where
\begin{eqnarray}
{\cal F}[X] &=& \frac{1}{\beta} \int_0^{\beta} d \tau \left\{
\frac{M}{2} \left(\frac{d X}{d \tau}\right)^2 + \frac{X^2(\tau)}{2} \right.
\nonumber
\\
&-& \left.
\frac{1}{4 \beta} X^2(\tau) \int_0^{\beta} d\tau' X^2(\tau')
\right.
\nonumber
\\
&+& \left. \frac{\eta}{4 \pi} \int_{-\infty}^{+\infty} d \tau'
\frac{(X(\tau)-X(\tau'))^2}{(\tau-\tau')^2}
\right\}
\label{fx}
\end{eqnarray}
is the replica free energy for $X(\tau)$. The coefficients in 
(\ref{fx}) are:
\begin{eqnarray}
M &=& \frac{E_1}{E_2} \, \, r^{-1}
\nonumber
\\
\eta &=& \frac{\gamma G_{\zeta}}{E_2} \, \, r^{-(1+\zeta/2)}
\label{meta2}
\end{eqnarray}
that can be associated with the ``particle'' mass and dissipation
coefficient, respectively. Here $G_{\zeta} = \int d^d q \, q^{-\zeta}
|\phi_q|^2$, $\phi_q$ is the Fourier transform of $\phi(z)$
and periodic boundary conditions are assumed: $X(\tau + \beta) = X(\tau)$.
Observe that ${\cal F}$ describes a Caldeira-Leggett action \cite{cl}
for the motion of a dissipative particle in a non-local, non-linear,
potential which, for slowly varying configurations of $X(\tau)$ reduces to
the potential of Fig.\ref{pot}.
Assuming that most of the time the field is in the equilibrium configuration
($X(\tau) = \pm 1$) the problem becomes equivalent to
the problem of a two-level system coupled to a dissipative
environment \cite{leggett}.

The amount of damping in the tunneling of the droplet can be
estimated by comparing the parameters in (\ref{meta2}).
We have weak damping when $\eta \ll \omega_0 M$ while for strong damping
one has $\eta \gg \omega_0 M$ ($\omega_0 = 2$ is the
undamped frequency of motion in our dimensionless units).
Notice that the crossover from
weak to strong dissipation occurs at $r = r_{c}$
where (using (\ref{meta2}))
\begin{eqnarray}
r_c \approx \left(\frac{\gamma G_{\zeta}}{2 E_1} \right)^{2/\zeta}
\nonumber
\end{eqnarray}
which indicates that close to the QCP ($r < r_c$) the droplet motion is highly
damped. The tunneling splitting between the two configurations $X = \pm 1$
is given by \cite{leggett}:
\begin{eqnarray}
\Delta = \omega_{cl} e^{-8 M \omega_{cl}}
\nonumber
\end{eqnarray}
where $\omega_{cl}$ is the classical frequency of oscillation. On the
one hand in the
weakly dissipative regime ($r>r_c$) one has
$\omega_{cl} \approx \omega_0$ and therefore the
tunneling splitting is:
\begin{eqnarray}
\Delta_{r>r_c}(r) \approx 2 e^{-a/r}
\label{deltaw}
\end{eqnarray}
where $a = 16 E_1/E_2$ is a universal constant.
On the other hand in the
strongly dissipative regime ($r<r_c$) we have
$\omega_{cl} \approx M \omega_0^2/\eta$ and therefore
\begin{eqnarray}
\Delta_{r<r_c}(r) \approx \frac{4 r^{\zeta/2}}{\gamma} e^{- b r^{\zeta}}
\label{deltas}
\end{eqnarray}
where $b = 32 E_1/(\gamma G_{\zeta})$
is a non-universal constant. For antiferromagnets ($\zeta=0$)
the tunneling splitting is constant close to the QCP while it vanishes
in the case of ferromagnets ($\zeta=1,2$). Another important parameter
is the Caldeira-Leggett dissipative coupling that is given by:
\begin{eqnarray}
\alpha(r) = \frac{2 \eta}{\pi} = \left(\frac{r_0}{r}\right)^{1+\zeta/2}
\label{alpha}
\end{eqnarray}
where we used (\ref{meta2}) and defined
\begin{eqnarray}
r_0 = \left(\frac{2 \gamma G_{\zeta}}{\pi E_2}\right)^{1/(1+ \zeta/2)} \, .
\label{r0}
\end{eqnarray}
Observe that the dissipative coupling diverges at the QCP. As is well-known
this problem has a characteristic crossover energy scale that can be associated
with the Kondo temperature, $T_K$, of an anisotropic Kondo impurity model
\cite{leggett}. This energy scale separates the region of strong and weak
coupling and for $\alpha < 1$ is given by:
\begin{eqnarray}
T_K(r) &\approx& W \left(\frac{\Delta(r)}{W}\right)^{\frac{1}{1-\alpha(r)}}
\label{tkdroplet}
\end{eqnarray}
where $W$ is a cut-off energy scale. For $\alpha > 1$ we have
$T_K = 0$ and the droplet is frozen. Notice that according to (\ref{alpha})
the freezing of the droplet occurs for $r<r_0$. This indicates that close
to the QCP droplets are frozen. Observe that $r_0$ given in (\ref{r0})
is a non-universal quantity which depends on parameters that cannot be
obtained in a continuum field theory. Depending on the microscopic
parameters the freezing of the droplets can occur in the region of
weak damping if $r_0 > r_c$ or strong damping if $r_0 < r_c$, affecting
the value of the tunneling splitting given in (\ref{deltaw}) and
(\ref{deltas}).
Thus $T_K(r)$ is finite and a continuous function of the distance
from the QCP in agreement with our previous analysis \cite{us2}.
This should be contrasted with the result obtained by MMS \cite{mms}
where $T_K = 0$ in all the parameter space.
In fact for $r \gg r_0$ the droplets become free to tunnel \cite{us1}.
The phase diagram is shown in Fig.\ref{diagram}.

\begin{figure}
\epsfysize3 cm
\hspace{0cm}
\epsfbox{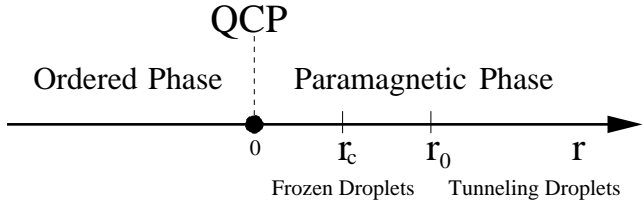}
\caption{Phase diagram as a function of the parameter $r$: the system
is ordered for $r<0$; droplets are frozen in the paramagnetic phase
for $0<r<r_0$ and can quantum tunnel for $r>r_0$.}
\label{diagram}
\end{figure}

Direct comparison of (\ref{tkdroplet}) with our previous
results in Ref.\cite{us2}
indicates that the two problems map into each other if $N
\propto \xi^2$ where $N$ is the number of spins in the cluster.
Another interesting consequence of our calculation is the
sharp contrast between the case of ferromagnetic and antiferromagnetic
droplets. For antiferromagnetic droples ($\zeta=0$) the dissipation
coefficient in (\ref{alpha}) scales with $\xi^2$ while
in the case of a clean ferromagnet ($\zeta=1$) we find $\alpha \propto
\xi^{3}$ indicating stronger damping. For a
ferromagnetic system with diffusive electrons ($\zeta=2$)
damping becomes even stronger with $\alpha \propto \xi^4$. This
indicates, in agreement with our previous analysis \cite{us2},
that dissipation is more important in the case of ferromagnetic
droplets than in the case of antiferromagnetic ones. Since the
great majority of systems studied experimentally are of the
antiferromagnetic type \cite{review} our results indicate that
dissipation does not freeze the droplets in a region of
the parameter space around the QCP.

A question that comes to mind is the reason for the
difference between our results
and the ones obtained by MMS \cite{mms}. We claim that our approaches
are valid in different regimes. A clear way to understand this difference
is to perform the same calculation for a Poisson distribution of
Dirac delta potentials \cite{claudio}. In this case the problem
is characterized by two physical parameters: the density of impurities
$\rho$ and the strength of the potential $V$. A universal Gaussian
distribution like the one discussed in this paper is obtained by
taking the limit of $\rho \to \infty$ and $V \to 0$ so that $u = \rho V^2$
is constant. On the other hand the single impurity limit studied by
MMS is obtained by letting $\rho \to 0$.

In summary, we have studied the problem of droplet formation and
dynamics close to a QCP of a disordered itinerant Ising magnet.
We find that the droplets have a finite Kondo temperature (associated with
their dissipative quantum dynamics) that varies continuously
with the distance from the QCP. We show that the characteristic Kondo
temperature of the droplet is finite except in a non-universal region
(dependent on microscopic details) close to the QCP.
These results are in agreement with our previous
analysis of quantum Griffiths-McCoy singularities in the disordered
Kondo lattice model and extends the validity of our results
to other strongly correlated systems.

We acknowledge invaluable discussions with
H. Castillo, C. Chamon, V. Dotsenko, A. J. Millis,
D.~Morr and J. Schmalian.

$^*$On leave from Department of Physics,
University of California, Riverside, CA 92521.

\end{document}